\documentclass[twocolumn,english,aps]{revtex4-2}
\usepackage{mathptmx}

\usepackage[T1]{fontenc}
\usepackage[utf8]{inputenc}
\usepackage{color}
\usepackage{babel}
\usepackage{array}
\usepackage{multirow}
\usepackage{amsbsy}
\usepackage{amstext}
\usepackage{graphicx}
\usepackage{textcomp}
\usepackage{hhline}
\usepackage{xcolor} 
\usepackage[unicode=true,pdfusetitle,
 bookmarks=true,bookmarksnumbered=false,bookmarksopen=false,
 breaklinks=false,pdfborder={0 0 0},pdfborderstyle={},backref=false,colorlinks=true]
 {hyperref}
\hypersetup{
 urlcolor=blue, citecolor=blue}

\makeatletter

\usepackage{braket}

\newcommand{\FeFam}{BaFe\textsubscript{2-x}M\textsubscript{x}As\textsubscript{2}}
\newcommand{\GX}{$\Gamma\text{X}$}
\newcommand{\GM}{$\Gamma\text{M}$}
\newcommand{\GZ}{$\Gamma\text{Z}$}
\newcommand{\dxy}{$d_{xy}$}
\newcommand{\dxz}{$d_{xz}$}
\newcommand{\dyz}{$d_{yz}$}
\newcommand{\dzz}{$d_{z^2}$}
\newcommand{\dxxyy}{$d_{x^2-y^2}$}

\makeatother

\begin{document}
\title{Orbital-Selective Band Structure Evolution in \FeFam{} (M = Cr, Co, Cu, Ru and Mn) Probed by Polarization-Dependent ARPES}
\author{K. R. Pakuszewski$^{1}$, M. R. Cantarino$^{2}$, I. Romanenko$^{1}$, A. P. Machado$^{3}$, M. M. Piva$^{4}$, G. S. Freitas$^{3}$, H. B. Pizzi$^{3}$, F. A. Garcia$^{5}$, P. G. Pagliuso$^{3}$ and C. Adriano$^{1}$}
\affiliation{$^{1}$Institut Quantique and Département de Physique, Université de Sherbrooke, 2500 boulevard de l’Université, Sherbrooke, Québec J1K 2R1, Canada}
\affiliation{$^{2}$European Synchrotron Radiation Facility, BP 220, F-38043 Grenoble Cedex, France}
\affiliation{$^{3}$Instituto de Física “Gleb Wataghin”, UNICAMP, 13083-859, Campinas-SP, Brazil}
\affiliation{$^{4}$Max Planck Institute for Chemical Physics of Solids, N\"othnitzer Str. 40, D-01187 Dresden, Germany}
\affiliation{$^{5}$Instituto de Física, Universidade de São Paulo, 05508-090 São Paulo, SP, Brazil}

\begin{abstract}

We present a systematic study of the evolution of the band structure in the Fe-based superconductor family \FeFam{} (M = Co, Cu, Cr, Ru and Mn) using polarization-dependent angle-resolved photoemission spectroscopy (ARPES). Low-substituted samples, with comparable spin-density wave transition temperatures ($T_\text{SDW}$), were chosen to enable controlled comparisons. The sizes of the central hole pockets ($\alpha$, $\beta$, and $\gamma$) remain largely unchanged across different substitutions, showing no clear correlation with either $T_\text{SDW}$ or the As height relative to the Fe planes. However, subtle trends are observed: a modest increase of the electron pocket with planar character that correlates with $T_\text{SDW}$ suppression, while its contraction appears linked to an increase in the As height relative to the Fe planes. Our results suggest that the suppression of $T_\text{SDW}$ is primarily driven by changes in the Fe–As bond length, with the effect being more pronounced in electronic states with planar character. These findings provide insight into the electronic structure of \FeFam{}.


\end{abstract}
\maketitle

\section{Introduction}

The discovery of superconductivity in fluorine-doped LaOFeAs by Kamihara \textit{et al}. \citep{kamihara_iron-based_2008} marked a groundbreaking advancement in the field of condensed matter physics, unveiling a new class of superconductors: the Fe-based superconductors (FeSC). This discovery ushered in a new era of research, sparking extensive investigations into the structural, electronic, and superconducting properties of these materials. The first family of compounds to be identified, known as the 1111 family, was quickly followed by the discovery of other families, most notably the 122 family, among which the parent compound BaFe$_{2}$As$_{2}$ (BFA)
\citep{ba122_rotter,ba122_canfield} is particularly well explored. The latter exhibits properties suitable for technological applications \citep{hosono_recent_2018}. A common structural feature across all these families is the presence of FeAs tetrahedra, which have been widely implicated in influencing the physical properties of these materials \citep{fe_based_hosono,fe_based_paglione,Stewart2011}.

Current hypotheses in the literature suggest that structural distortions within the FeAs tetrahedra play a critical role in modifying the physical properties of these materials, potentially driving the system into a superconducting state. Specifically, it has been proposed that an optimal distance between the As atoms and the Fe plane, along with a specific Fe–As–Fe bond angle, are crucial factors favoring the emergence of superconductivity \citep{Yang2014,Rotter2008,Hardy2009}. Lee \textit{et al.}\citep{Lee2008} were the first to report that an Fe–As–Fe bond angle of approximately $109.5^\circ$ favors high superconducting critical temperature ($T_\text{SC}$) values. Later, Mizuguchi \textit{et al.}\citep{Mizuguchi2010} showed that an optimal As height relative to the Fe planes of approximately $1.38,\text{\AA}$ also promotes high $T_\text{SC}$.

This sensitivity of the physical properties to structural distortions of the FeAs tetrahedra is intimately connected to changes in the orbital contributions to the bands forming the Fermi surface (FS) in these materials. In particular, spectroscopic and theoretical studies have shown that the density of states (DOS) of the FeSCs near the Fermi level ($E_F$) is almost entirely derived from electrons in the five Fe $3d$ orbitals \citep{ba122_dos_nekrasov,ba122_dos_li,ba122_dos_sen,jishi2010,lu2009,shim2009}. Due to the crystal-field environment of the FeAs tetrahedra, the degeneracy of these five $3d$ orbitals is lifted into three nearly degenerate $t_{2g}$ states (\dxy{}, \dxz{}, and \dyz{}) and two $e_g$ states (\dxxyy{} and \dzz{}). Consequently, even small perturbations to the tetrahedral geometry can induce significant changes in the orbital makeup of the FS and, therefore, modify the physical properties of the FeSCs.

Subsequent studies have proposed that an enhanced population of planar-electron orbitals (primarily \dxy{} and \dxxyy{}) at the FS may play a crucial role in promoting high-temperature superconductivity (HTSC) \citep{Suzuki2014,fe_based_hosono}. The sensitivity of HTSC to the details of the electronic structure reflects its close connection to charge and spin fluctuations, which remain central to our current understanding of the FeSCs. Various theoretical approaches emphasize either the itinerant or localized character of these fluctuations \citep{fernandes_iron_2022}, leading to different interpretations of how chemical substitutions influence the electronic structure and related physical properties.

Several theoretical and experimental investigations have attempted to clarify how the Fe $3d$ electrons organize into the bands that constitute the FS of the FeSCs. Yin \textit{et al.} \citep{ba122_fs_the_yin}, for example, demonstrated that in the BFA parent compound the FS comprises five pockets: three hole pockets centered at the $\Gamma$ point of the Brillouin zone (BZ) and two electron pockets at the X point. The FS obtained using density functional theory combined with dynamical mean-field theory (DFT+DMFT) closely matches the FS observed in our ARPES measurements, indicating that electronic correlations play an important role in the \FeFam{} family. These findings are consistent with other theoretical \citep{ba122_dos_fs_roe,ba122_dos_fs_graser,Wang2015} and ARPES \citep{ba122_exp_fs_pfau,ba122_exp_fs_Evtushinsky,Mansart2011} studies, which report similar FS shape and orbital contribution in BFA and its substituted compounds.

Concerning the orbital contributions to the electron and hole pockets, theoretical works such as those by Yin \textit{et al.} \citep{ba122_fs_the_yin} and van Roekeghem \textit{et al.} \citep{ba122_dos_fs_roe} generally suggest that the FS is dominated by $t_{2g}$ orbitals, whereas the $e_g$ orbitals lie well below $E_F$ or contribute only at specific regions of the BZ, such as near the Z point. Similar conclusions were reached by Pfau \textit{et al.} \citep{ba122_exp_fs_pfau} in their ARPES study of twinned and detwinned BFA, where nematicity was shown to induce a splitting between the \dxz{} and \dyz{} orbitals and to significantly renormalize the low-energy band structure.

On the other hand, several experimental results point to additional complexity. Zhang \textit{et al.} \citep{Zhang2011}, studying optimally Co-substituted BFA, reported that the outer hole pocket exhibits a mixed \dxy{} and \dxxyy{} orbital character. They also found that both inner hole bands lie below $E_F$ and thus do not contribute to the FS, leaving only the outer hole band crossing the Fermi level. Similar orbital assignments were reported by Evtushinsky \textit{et al.} \citep{ba122_exp_fs_Evtushinsky} for K-substituted BFA, with the main difference being that in this case both inner hole pockets contribute to the FS. Our results are consistent with these previous studies regarding the orbital character of the hole and electron pockets, while also revealing substitution-dependent energy shifts of the inner hole bands.

In this work, we focus on ARPES measurements of the \FeFam{} (M = Cr, Co, Cu, Ru, and Mn) electronic band structures to investigate the evolution of the hole and electron pockets as a function of chemical substitution. Here the samples will be denoted as BFA, Cr-BFA, Co-BFA, Cu-BFA, Ru-BFA and Mn-BFA for the parent and substituted compounds. By selecting compounds sharing approximately the same spin-density wave (SDW) transition temperature ($T_\text{SDW}$), we establish a control parameter that allows us to systematically explore the relationship between macroscopic physical properties, such as $T_\text{SDW}$ or structural parameters, and the size and shape of the bands forming the FS.   Our primary objective is to determine whether the physical properties of these materials, and consequently their band structure, are solely influenced by structural distortions of the FeAs tetrahedra, as suggested in the literature, or whether additional parameters must be considered in shaping the electronic structure.

In addition, we examine the evolution of the scattering rate as a function of the binding energy extracted from ARPES spectra across the entire \FeFam{} series. A particular focus is given to its energy dependence, as our previous studies on Mn- and Cr-substituted compounds \citep{marli_Mn,marli_Cr}, as well as work by Fink \textit{et al.} \citep{fink_paper}, suggest a linear behavior characteristic of a marginal Fermi liquid. Our goal was to determine whether this behavior persists across different elements and to what extent it reflects the underlying electronic correlations in these systems.

In the low-substitution regime, we were not able to find a clear correlation between the size of the central hole pockets and either the $T_\text{SDW}$, the As height structural parameter, or the electron/hole character of the substitution. However, we observed a systematic trend in the size of the electron pockets derived from planar orbitals (\dxy{} and \dxxyy{}), which show a moderate enlargement correlated with decreasing $T_\text{SDW}$ (as shown in Figure \ref{fig_size_center}(c)). 
	
These findings, together with previous EXAFS studies \citep{bittar_exafs} that report a reduced Fe–As bond length upon substitution, suggest that bands with planar orbital character are particularly sensitive to local structural changes and may play a key role in the evolution of magnetism and superconductivity in lightly substituted \FeFam{}. To test this hypothesis, we compared the size of the planar electron pocket with the As height extracted from our x-ray diffraction measurements and found a clear correlation: an increase in As height corresponds to a shrinking of the planar electron pocket.

\section{Material and Methods}

\begin{figure}[ht]
\centering
\includegraphics[width=\columnwidth]{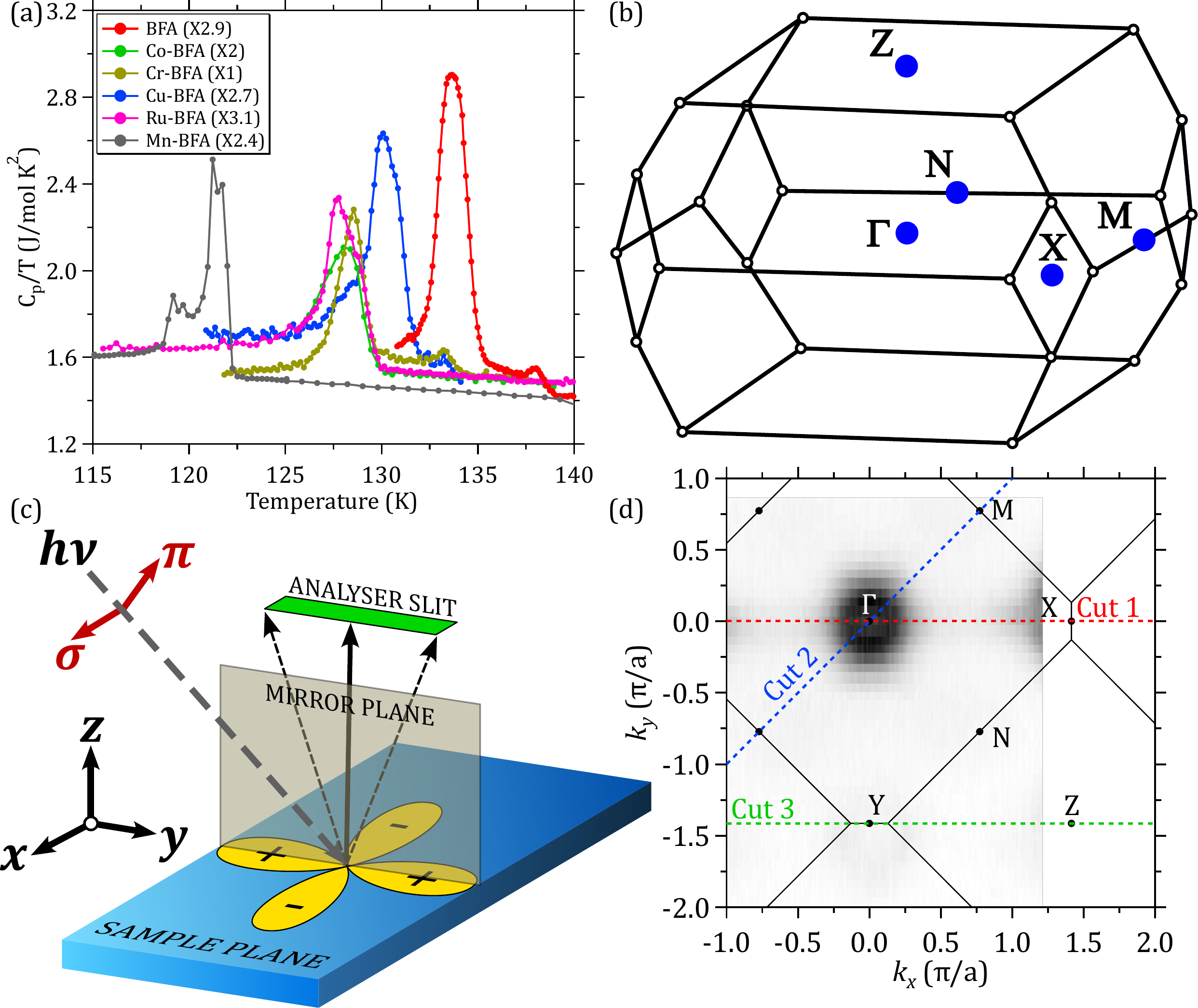}
\caption{Overview of the macroscopic measurements and ARPES experimental configuration for the \FeFam{} family. (a) Heat capacity as a function of temperature, highlighting the magnetic transition temperature $T_\textrm{SDW}$. The X plus a number means that the y scale was multiplied by this quantity. (b) Tetragonal Brillouin zone (BZ) of the BFA parent compound (space group I4/mmm). (c) Schematic of the polarization-dependent ARPES experimental setup. (d) FS of the parent compound measured along the \GX{} direction with $\sigma$ polarization, indicating three high-symmetry cuts (cut 1, cut 2 and cut 3).\label{fig_opening}}

\end{figure}

The \FeFam{} (M = Co, Cu, Cr, Ru, and Mn) samples were synthesized using the In-flux technique, as reported in previous literature \citep{garitezi_synthesis_2013}. These samples were carefully selected from the low-substituted region, where the SDW transition temperature ($T_\text{SDW}$) is approximately $127\,\text{K}$. This choice ensures a consistent control parameter across all compounds. The $T_\text{SDW}$ magnetic transition temperature was verified through heat capacity measurements, which are depicted in Figure \ref{fig_opening}(a). These measurements reveal a distinct peak in the $C_p/T$ curves, confirming the transition.

For comparative purposes, the BFA parent compound was also measured under the same conditions. Polarization-dependent Angle-Resolved Photoemission Spectroscopy (ARPES) measurements were conducted at the Quantum Materials Spectroscopy Centre (QMSC) beamline located at the Canadian Light Source (CLS). The samples were analyzed using a photon energy of $48\,\text{eV}$, which corresponds to a $k_z$ value near the $\Gamma$ point of the BZ, as illustrated in Figure \ref{fig_opening}(b). Measurements were taken along the high-symmetry directions \GX{} and \GM{} of the BZ, or parallel to them, utilizing both linear vertical (LV) and linear horizontal (LH) polarizations.

All samples were measured above $T_\text{SDW}$, in the paramagnetic state, at a temperature close to $150\,$K. Data were collected using a Scienta-Omicron R4000 electron analyzer, operating in an ultra-high vacuum environment with a pressure better than $10^{-11}\,$mbar. Additional $k_z$ dispersion measurements were carried out at the Bessy II synchrotron on the One-Square beamline, covering an energy range between $34\,$eV and $48\,$eV. These measurements were performed below $T_\text{SDW}$, at a temperature of approximately $20\,$K.

The orbital sensitivity of the Fe $3d$ electrons to light polarization provides a powerful tool for probing the contributions of different orbitals to the bands that form the FS. Figure \ref{fig_opening}(c) illustrates the experimental setup for a polarization-dependent ARPES experiment. Depending on the parity of the orbitals relative to the mirror plane defined in Figure \ref{fig_opening}(c), the ARPES signal for specific electrons can be completely nullified under certain conditions. Consequently, electrons from particular Fe $3d$ orbitals may not be detected, depending on the crystal orientation (\GX{} or \GM{} direction) and the polarization of the light. When the light polarization is parallel to the mirror plane, it is referred to as $\pi$ polarization (corresponding to LH polarization in our beamline setup), and when it is perpendicular to the mirror plane, it is termed $\sigma$ polarization (LV polarization in our beamline setup). As summarized in Table \ref{tab_polari}, only electrons originating from orbitals marked with an X will be detectable under these experimental conditions \citep{Damascelli2003}.

\begin{table}[ht]
\begin{centering}
\caption{Orbital selection rules table as a function of the experimental geometry. A \textquotedblleft X\textquotedblright{} symbol denotes that the band can be observed for the indicated polarization and experimental geometry. The parity of each orbital is based on the setup illustrated in Figure \ref{fig_opening}-c. \label{tab_polari}}
\par\end{centering}
\centering{}%
\begin{tabular}{ccccccc} 
\hhline{=======}
\multirow{2}{*}{Direction}          & \multirow{2}{*}{Polarization} & \multicolumn{5}{c}{Fe $3d$ Orbitals}                                                                         \\ 
\cline{3-7}
                                    &                               & \hspace{0.014\textwidth}\dxz\hspace{0.014\textwidth} & \hspace{0.014\textwidth}\dxxyy\hspace{0.014\textwidth} & \hspace{0.014\textwidth}\dzz\hspace{0.014\textwidth} & \hspace{0.014\textwidth}\dyz\hspace{0.014\textwidth} & \hspace{0.014\textwidth}\dxy\hspace{0.014\textwidth}  \\ 
\hline
\multirow{2}{*}{\GX} & $\pi$        & X                   & X                     & X                   &                     &                      \\
                                    & $\sigma$     &                     &                       &                     & X                   & X                    \\
\hline
\multirow{2}{*}{\GM} & $\pi$        & X                   &                       & X                   & X                   & X                    \\
                                    & $\sigma$     & X                   & X                     &                     & X                   &                      \\
\hline\hline
\end{tabular}
\end{table}

\section{Results and Discussion}

In Figure \ref{fig_opening}(d), we show the FS of the BFA parent compound measured along the \GX{} direction with $\sigma$ polarization. All substituted samples exhibit similar FSs, characterized by two or three hole pockets centered at the $\Gamma$ point and two electron pockets located at the X/Y points. These features are consistent with previous reports on the BaFe$_2$As$_2$ family \citep{ba122_dos_fs_graser,Mansart2011,Wang2015,ba122_dos_fs_roe,ba122_exp_fs_pfau,ba122_exp_fs_Evtushinsky,ba122_fs_the_yin}, and align with our earlier results for Mn- and Cr-substituted compounds \citep{marli_Mn, marli_Cr}. The reason why some studies report two hole pockets while others observe three is likely due to the behavior of the inner hole band ($\beta$ band), which reaches a maximum near the Fermi level ($E_F$); in some compounds it crosses $E_F$, while in others, it does not. 

Our analysis primarily focuses on the hole bands around the $\Gamma$ point and the electron bands around the X/Y points of the BZ. For the hole bands, measurements were performed along the high-symmetry directions \GX{} and \GM{} (cuts 1 and 2 in figure \ref{fig_opening}(d)). For the electron bands, measurements were carried out along the \GX{} and YZ directions (cuts 1 and 3 in figure \ref{fig_opening}(d)), with cut 1 centered at either the $\Gamma$ or X point to probe the hole or electron pocket, respectively. These measurements aim to track the evolution of the band structure across different elemental substitutions, all with similar $T_\text{SDW}$, with particular attention to changes in the size of the hole and electron pockets within the BFA family. To this end, we acquired several high-statistics bandmaps (BMs) for each compound along the indicated cuts, enabling a detailed analysis of the bands forming the FS pockets. In the following figures, we present a comprehensive analysis of these BMs.

To determine the position of each band, we first applied a second derivative or curvature analysis to the raw BMs. This procedure enhances the visibility of band features and provides more accurate initial estimates for fitting the original data. A detailed description of this method can be found in the Supplemental Material \citep{ref_sup}.

It is important to note that both the second derivative and curvature techniques offer different advantages and disadvantages. The second derivative method often reveals weaker bands, but it tends to broaden both weak and strong bands. In contrast, the curvature method produces narrower bands but may suppress weaker bands. Consequently, we chose the most appropriate technique for each BM individually. In the processed data, the original spectral peaks become valleys, and the minima indicate the approximate band positions, which were then used as initial guesses for subsequent fits to the raw data. It is important to note that the fitting procedure was applied only to extract the band positions of the hole bands. For the electron bands, we relied solely on the initial positions obtained from second derivative or curvature analysis. This choice was made because fitting the electron bands proved challenging, due to the high density of closely spaced bands within the electron pockets.

\begin{figure}[ht]
\centering
\includegraphics[width=\columnwidth]{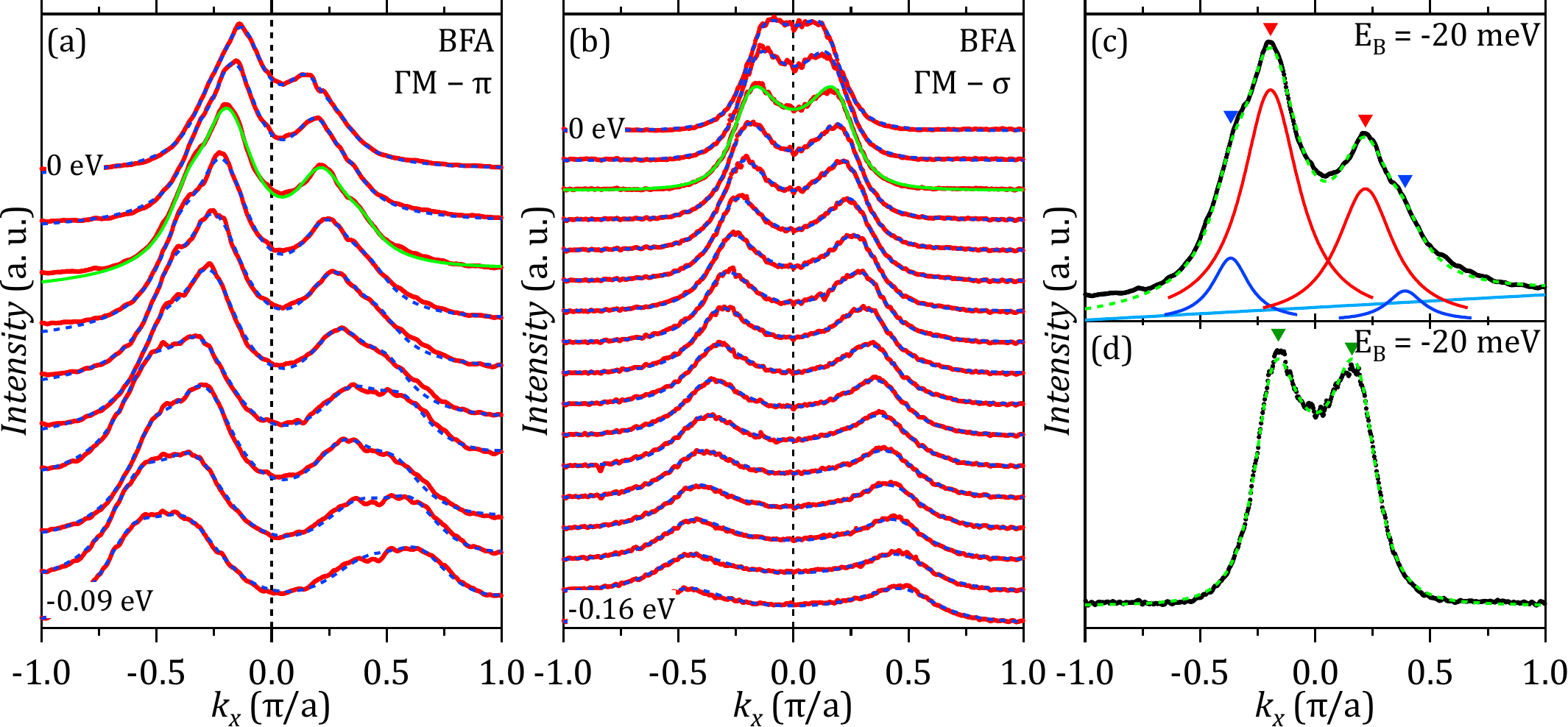}
\caption{ARPES spectral function analysis. MDC waterfall plots of the BFA compound along the \GM{} direction measured with (a) $\pi$ and (b) $\sigma$ polarization. Red circles represent the MDC data, while blue dashed lines indicate the corresponding fits. Panels (c) and (d) display the MDC fits highlighted by green solid lines in panels (a) and (b), respectively. Here, black circles represent the experimental MDC data, green dashed lines show the overall fit, and light blue lines correspond to the background. In panel (c), individual Lorentzian peaks used in the fit are also shown, with blue and red curves corresponding to the $\gamma$ and $\alpha$ bands, respectively.\label{fig_waterfall}}

\end{figure}

Figure \ref{fig_waterfall}(a) and (b) presents momentum distribution curves (MDCs) in a waterfall plot with their specific fittings for the BFA parent compound. Two different fitting approaches were employed, both based on MDCs fitting, which are horizontal cuts at constant energy in the BMs. In the first approach (Figure \ref{fig_waterfall}(c)), each band is modeled as a pair of Lorentzian peaks in the MDC spectra, with mirroed positions and equal widths. This method, widely used in ARPES analysis \citep{Damascelli2003,Matsuyama2013,Avigo2017,Zhang2013}, approximates the spectral function as a sum of Lorentzian peaks plus a background term (typically constant or linear). Due to the large number of fitting parameters involved, our initial fits fixed the peak positions to those obtained from the second derivative/curvature analysis. Once other parameters were determined, we performed a second fitting step in which the peak positions were allowed to vary while keeping some of the previously fitted parameters fixed.

Figures \ref{fig_waterfall}(a) and (c) show the fitting results for data along the \GM{} direction with $\pi$ polarization, using the Lorentzian peak model. The same method was also applied to data along the \GX{} direction for both polarizations (not shown). In Figure \ref{fig_waterfall}(a), several MDCs from binding energies $E_B = -0.09\,$eV to $0\,$eV (red points) are plotted along with their respective fits (blue dashed lines). A representative fit (full green line) is highlighted in Figure \ref{fig_waterfall}-c, where the individual Lorentzian peaks (blue and red lines), linear background (light blue line), and final fit (green dashed line) are shown. The positions of the individual peaks mark the band positions in the MDCs.

In the second approach (Figure \ref{fig_waterfall}(d)), we attempted to directly fit the ARPES spectral function to the data. The spectral function is defined as:

\begin{equation}
\label{eq_arpes_spectra}
A(E,k,T)=\frac{1}{\pi}Z(E,k,T)\frac{\frac{\Gamma(E,k,T)}{2}}{\left[ E-\epsilon^*_k \right]^2+\left[ \frac{\Gamma(E,k,T)}{2} \right]^2}
\end{equation}

\noindent where $\Gamma(E,k,T)$ represents the scattering rate and $Z(E,k,T)$ is the renormalization function, both of which depend on energy, momentum, and temperature. The term $\epsilon^*_k$ denotes the renormalized dispersion \citep{Damascelli2003,scatrate_slope_fink}. For the dataset measured along \GM{} with $\sigma$ polarization, only one band is visible, and it exhibits a parabolic dispersion. Therefore, we approximated $\epsilon^*_k$ using a simple parabolic model:

\begin{equation}
\epsilon^*_k = \epsilon_0 - ax^2
\end{equation}

Furthermore, we assumed that $\Gamma(E,k,T)$ and $Z(E,k,T)$ are approximately momentum-independent. And since we are analyzing MDCs at constant energy and temperature, both quantities were treated as constant fitting parameters. Under this assumption, the only momentum-dependent term in Equation \ref{eq_arpes_spectra} is the renormalized dispersion $\epsilon^*_k$. This simplification allows the band position to be extracted from the roots of the quadratic equation $E - \epsilon^*_k = 0$. Figures \ref{fig_waterfall}(b) and (d) display the fitting results using this second method. Further details on both fitting approaches are available in the Supplemental Material \citep{ref_sup}.

\begin{figure}[ht]
\centering
\includegraphics[width=\columnwidth]{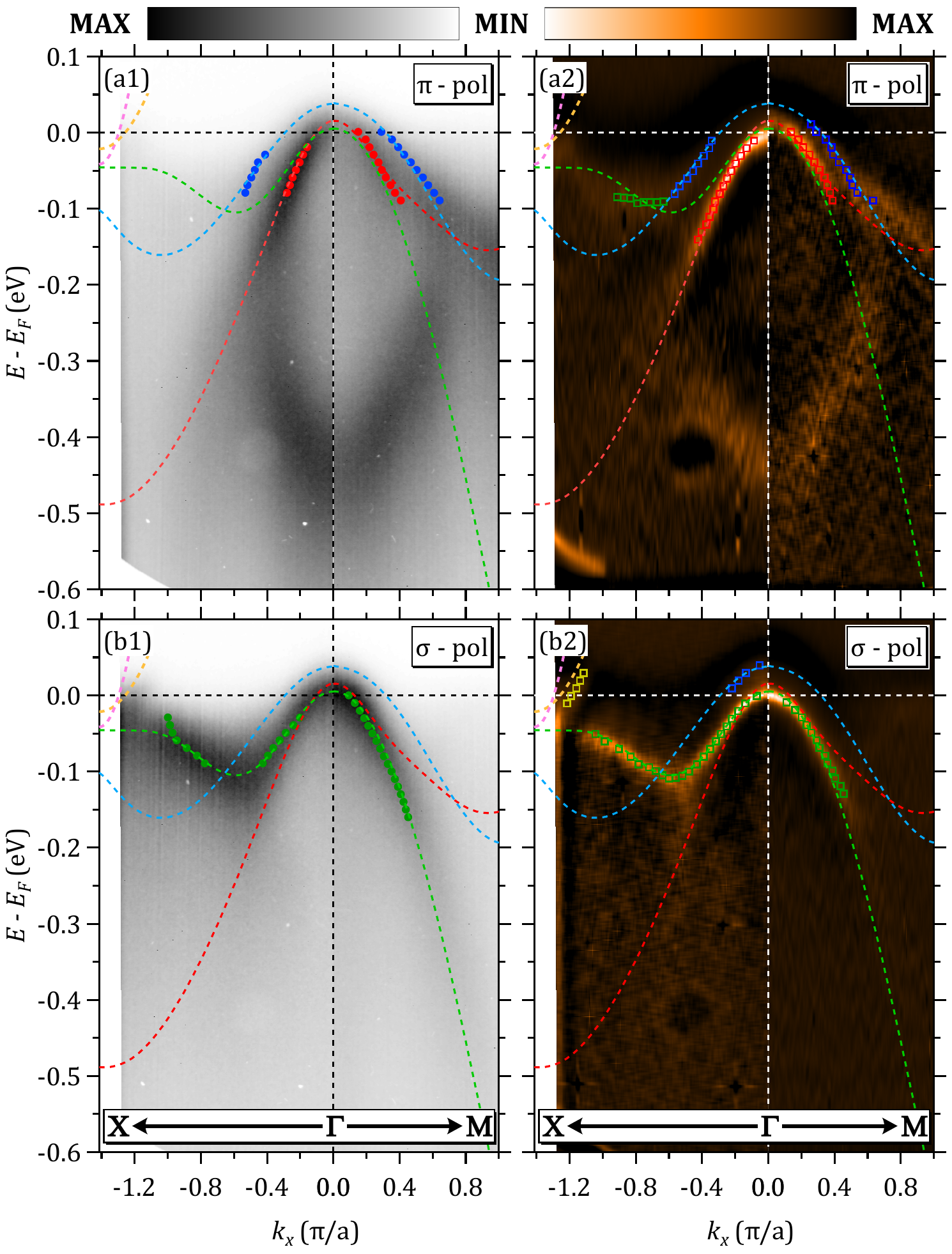}
\caption{Summary of the ARPES measurements for the BFA parent compound. The upper panels (a1 and a2) correspond to measurements performed with $\pi$ polarization, while the lower panels (b1 and b2) correspond to those taken with $\sigma$ polarization. Left panels (a1 and b1) show the raw BMs, and right panels (a2 and b2) display the corresponding curvature BMs. Dashed lines serve as guide-to-eye to indicate the band dispersions. The colors red, green, and blue denote the $\alpha$, $\beta$, and $\gamma$ bands, respectively. Furthermore, at the X corner, the two electron bands $\delta$ (pink) and $\eta$ (orange) are indicated with dashed lines. The open squares denotes the points extracted via fitting or second derivative/curvature analysis.
 \label{fig_center_bms}}

\end{figure}

Based on the extracted band positions, we constructed a schematic model of the band dispersions, shown as dashed lines in Figure \ref{fig_center_bms}. These schematics serve as guide-to-eye to the experimental band structure, following the trends extracted from both the second derivative/curvature analysis and the fitting procedures. They also reflect the general features observed in previous experimental \citep{Mansart2011,Zhang2011,ba122_exp_fs_pfau,ba122_exp_fs_Evtushinsky} and theoretical works \citep{ba122_dos_fs_graser,ba122_fs_the_yin,ba122_dos_fs_roe}.

Figure \ref{fig_center_bms} presents the raw BMs (left panels) alongside their corresponding curvature BMs (right panels) for the BFA compound in the four possible experimental configurations: along the \GX{} and \GM{} directions, with $\pi$ and $\sigma$ polarizations. Each band of interest is labeled using distinct colors and names: red for the $\alpha$ band, green for the $\beta$ band, blue for the $\gamma$ band, pink for the $\delta$ band and orange for the $\eta$ band.

An additional band, not marked with a dashed line, appears only under $\pi$ polarization with its band bottom located around $-450\,$meV. Theoretical calculations \citep{ba122_dos_fs_roe,ba122_dos_fs_graser} suggest that this band has a \dzz{} orbital character, which is consistent with the orbital selection rules presented in Table \ref{tab_polari}, where the \dzz{} orbital can only be detected under $\pi$ polarization.

\begin{figure}[ht]
\centering
\includegraphics[width=\columnwidth]{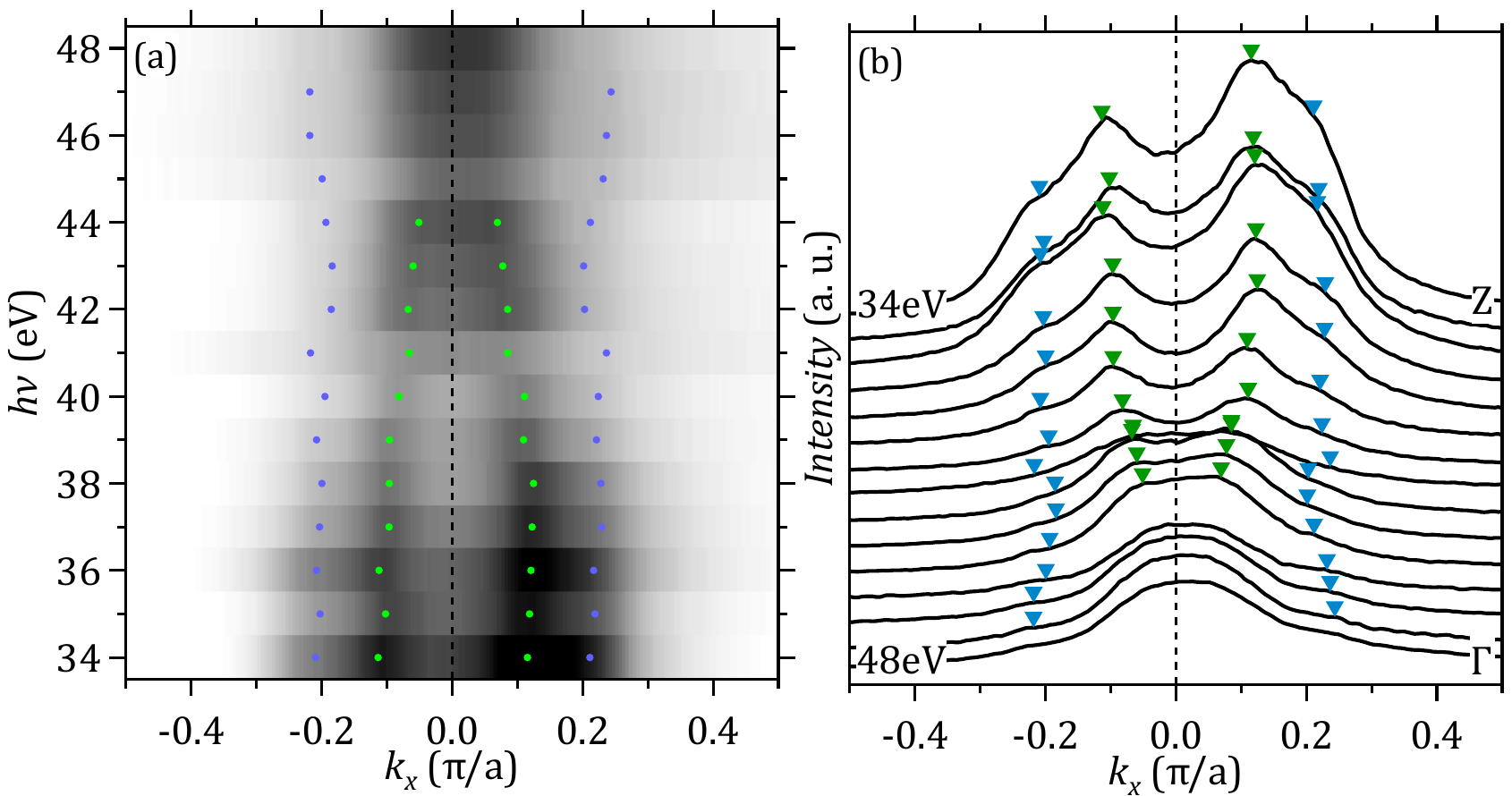}
\caption{$k_z$-dispersion measurements of Cr-BFA taken along the \GX{} direction using $\pi$ polarization. (a) Fermi surface map as a function of excitation energy ($h\nu$) and in-plane momentum ($k_x$). (b) MDC waterfall plot corresponding to the data in panel (a). Green and blue circles/triangles indicate the extracted band positions, obtained from Lorentzian fits to the MDCs.\label{fig_kz_disp}}

\end{figure}

Turning our attention to the central bands at the $\Gamma$ point and near $E_F$, the $\alpha$ band is observed under $\pi$ polarization in both crystallographic directions. Along the \GM{} direction, it appears slightly degenerate with the $\beta$ band, which is also visible under $\sigma$ polarization at the same momentum position of $\beta$ band (not marked on the graph). Several theoretical and experimental studies \citep{Zhang2011,ba122_exp_fs_pfau,ba122_dos_fs_graser,ba122_dos_fs_roe} support this degeneracy, suggesting that the $\alpha$ and $\beta$ bands share similar contributions from the \dxz{} and \dyz{} orbitals along \GM{} direction. Most theoretical works propose a dominant \dxz{} character with a mixed \dyz{} contribution for the $\alpha$ band along the \GM{} direction, as well as a parabolic dispersion in both directions, though the band reaches the X and M points at different binding energies. Our observations for the $\alpha$ band are in good agreement with these predictions and are consistent with the orbital selection rules summarized in Table \ref{tab_polari}. We therefore conclude that this band exhibits a strong \dxz{} character along the \GX{} direction, and a mixed \dxz{} and \dyz{} character along \GM{}.

\begin{figure*}[ht]
\centering
\includegraphics[width=\textwidth]{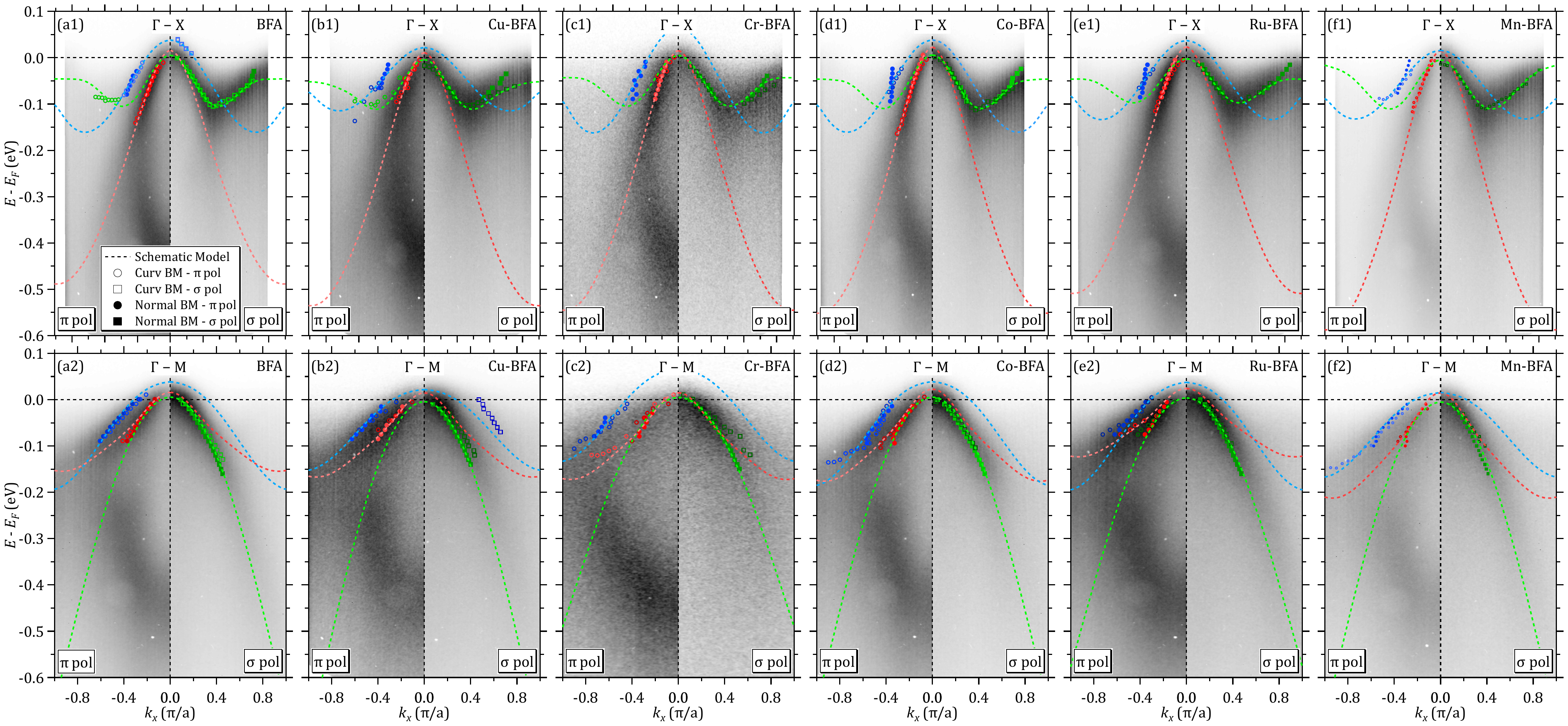}
\caption{Summary of the hole band dispersion analysis for the \FeFam{} family. Upper panels (a1 to f1) show data taken along the \GX{} direction, while lower panels (a2 to f2) show data taken along the \GM{} direction, both using $\pi$ and $\sigma$ polarizations. Circles represent data obtained with $\pi$ polarization, and squares with $\sigma$ polarization. Open markers (circles or squares) correspond to points extracted from curvature or second-derivative BMs, while filled markers indicate values obtained from MDC fitting of the raw BMs. Dashed lines are guide-to-eye of the band dispersions for each compound.\label{fig_all_central_bands}}

\end{figure*}

The $\beta$ band is observed under $\sigma$ polarization along both directions, and under $\pi$ polarization it appears degenerate with the $\alpha$ band exclusively along the \GM{} direction (not marked on the graph). This band shows a parabolic dispersion near the $\Gamma$ point but bends upward midway between $\Gamma$ and X. Along the \GM{} direction, the $\beta$ band maintains a parabolic shape throughout. Based on Table \ref{tab_polari}, we infer a strong \dyz{} orbital character with a \dxz{} admixture along \GM{}, consistent with previous theoretical and experimental reports on this compound.

The $\gamma$ band is more prominently observed along the \GM{} direction under $\pi$ polarization, while it appears with a much weaker intensity along the \GX{} direction and $\sigma$ polarization, where it is often overshadowed by the stronger $\beta$ band signal. This information suggests a mixing of even and odd orbital characters. As most theoretical studies propose a two-dimensional nature for this band, we attribute it primarily to contributions from the \dxxyy{} and \dxy{} orbitals. Moreover, in our fitting analysis, the $\gamma$ band consistently exhibits a broader linewidth and weaker spectral weight compared to the $\alpha$ and $\beta$ bands. This further supports the orbital character interpretation, as multiple studies have shown that electrons originating from the \dxy{} orbital typically exhibit weaker spectral weight and broader linewidths in ARPES measurements at high temperatures \citep{Sobota2021,Yi2013,Yi2015,Miao2016}. Notably, Yi \textit{et al.} \citep{Yi2015} demonstrated that in iron chalcogenides, bands dominated by \dxy{} electrons undergo significant spectral weight renormalization with increasing temperature, marked by a pronounced suppression of spectral weight.

To confirm the two-dimensional (2D) nature of the $\gamma$ band and the three-dimensional (3D) nature of the $\beta$ band, and consequently their respective orbital contributions, we performed $k_z$ dispersion measurements on the Cr-BFA compound along the central bands. As previously discussed, bands with predominant \dxy{} and/or \dxxyy{} orbital character are expected to exhibit a more 2D nature, and therefore should show minimal dispersion along $k_z$. In contrast, bands with dominant \dxz{}, \dyz{}, and \dzz{} orbital character are typically more 3D and exhibit significant $k_z$-dependent changes \citep{ba122_dos_fs_graser,Zhang2011,Zhang2010}.

Figure \ref{fig_kz_disp}(a) presents the $k_z$ dispersion measurements of the Cr-BFA compound along the \GZ{} direction under $\pi$ polarization. The $\gamma$ band (blue circles and triangles) shows only weak dispersion from $\Gamma$ to Z, consistent with a quasi-2D character. Figure \ref{fig_kz_disp}(b) displays the corresponding MDCs from panel \ref{fig_kz_disp}(a), where the inverted triangles mark the peak positions. The MDCs were fitted with four Lorentzian peaks to extract the band positions.

In contrast, the $\beta$ band (green circles and triangles) exhibits a much stronger $k_z$ dispersion, vanishing from the FS near the $\Gamma$ point. This behavior is characteristic of a three-dimensional band and supports the assignment of the $\beta$ band as having strong \dxz{}, \dyz{}, or \dzz{} orbital contributions. These observations further confirm the orbital character of both the $\beta$ and $\gamma$ bands. Similar behavior was reported by Zhang \textit{et al.} \citep{Zhang2011} for the $\alpha$ band, which also shows a highly 3D dispersion along $k_z$.

Having established the band structure analysis for the BFA parent compound, we now extend the same methodology to the substituted samples. This allows for a direct comparison across the entire \FeFam{} series. Figure \ref{fig_all_central_bands} summarizes the extracted band positions of the hole pockets for all compounds studied in this work.

In the figure \ref{fig_all_central_bands}, filled circles and squares represent band positions obtained by fitting the MDCs of the raw data. Open circles and squares correspond to positions determined from the minima of the MDCs in the curvature or second derivative data. Data measured with $\pi$ polarization are shown as circles (right side of the graphics), while data from $\sigma$ polarization are shown as squares (left side of the graphics). 

The upper panels display measurements along the \GX{} direction, and the lower panels along the \GM{} direction. The dashed lines represent our band model and are intended to closely follow the shape of the extracted points from both methods (minima and fitting). These lines serve as a guide-to-eye to the overall band dispersion trends across the different compounds.

\begin{figure}[ht]
\centering
\includegraphics[width=\columnwidth]{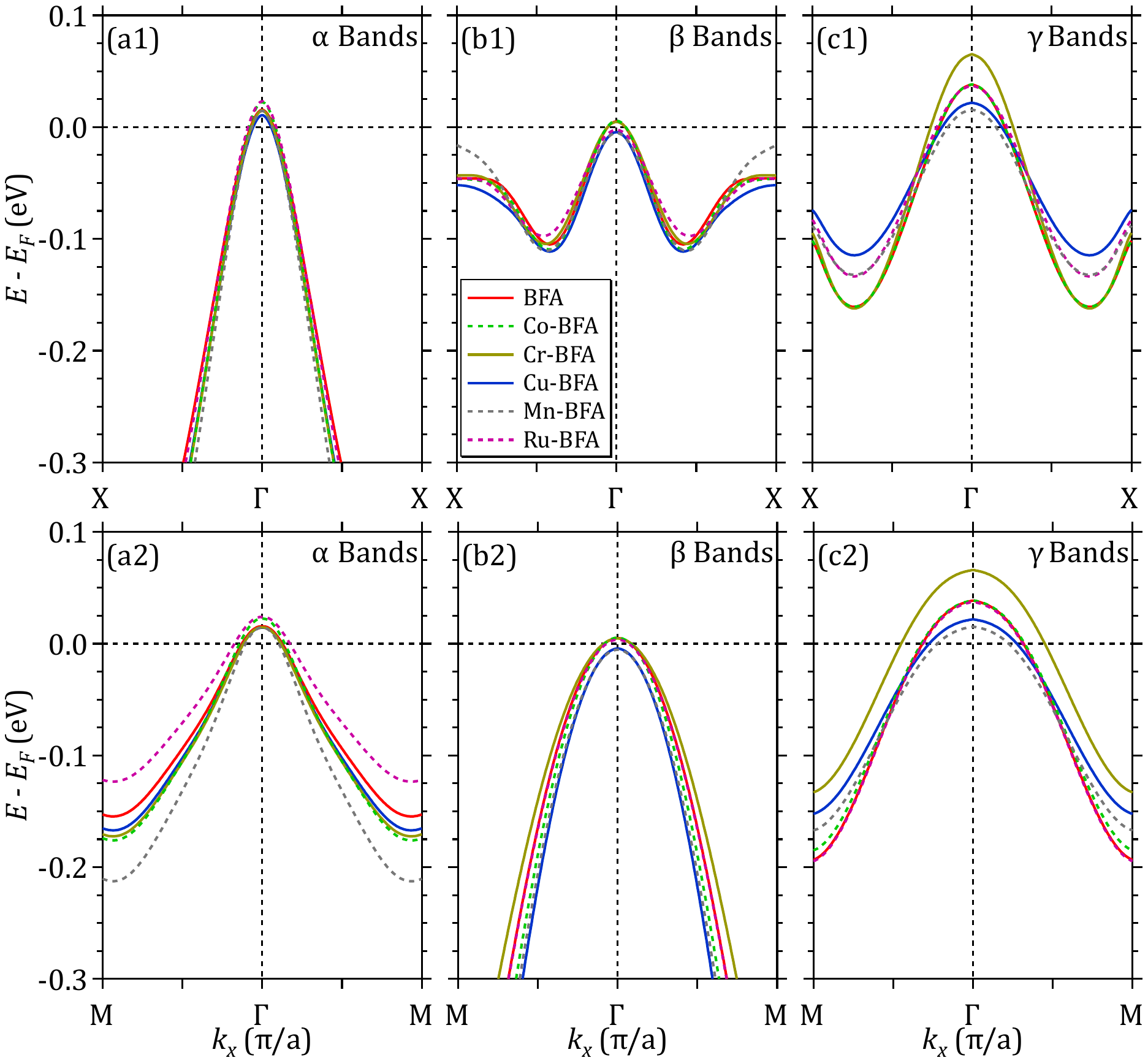}
\caption{Comparison of the hole bands across the \FeFam{} family. Upper panels (a1 to a3) show the schematic dispersions of the three hole bands measured along the \GX{} direction: (a1) $\alpha$, (a2) $\beta$, and (a3) $\gamma$. Lower panels (b1 to b3) present the same bands measured along the \GM{} direction: (b1) $\alpha$, (b2) $\beta$, and (b3) $\gamma$. \label{fig_comp_center_bands}}

\end{figure}

Figure \ref{fig_comp_center_bands} compares the hole bands of all measured compounds. The upper panels display the bands measured along the \GX{} direction, while the lower panels show the bands along the \GM{} direction. From left to right, the panels correspond to the $\alpha$, $\beta$, and $\gamma$ bands, which together form the hole pockets centered at $\Gamma$ in the FS.

As observed, the $\alpha$ and $\beta$ bands along the \GX{} direction (Figures \ref{fig_comp_center_bands}(a1) and (a2)) do not exhibit significant changes in their shapes across different compounds. Furthermore, the $\beta$ band only crosses the Fermi level ($E_F$) in some of the studied samples. However, due to experimental uncertainties in determining the exact $E_F$ position, it is possible that the $\beta$ band may cross $E_F$ in all compounds - or in none.

In contrast, the $\alpha$ band consistently crosses $E_F$ in all doped samples, though with a slight shift in the crossing point depending on the compound. Among them, Ru-BFA sample exhibits the largest $\alpha$ pocket. A similar trend is observed for the $\alpha$ and $\beta$ bands along the \GM{} direction (Figures \ref{fig_comp_center_bands}(b1) and (b2)), where the band shapes remain largely unchanged, but the $\alpha$ band shows a slight upward energy shift. Again, Ru-BFA compound presents the largest pocket in this direction. Yin \textit{et al.} \citep{ba122_fs_the_yin} suggest that these two bands form superellipse-like FS sheets (rounded squares), with the vertex of the inner pocket pointing toward the M direction and that of the outer pocket pointing toward the X direction. Both pockets exhibit a degeneracy along the \GM{} direction, where the vertex of the inner pocket touch the middle of the outer pocket, which is consistent with experimental observations.

We now turn our attention to the central $\gamma$ band (Figures \ref{fig_comp_center_bands}(a3) and (b3)). In comparison with the $\alpha$ and $\beta$ bands, this band show greater variation of the pocket size at $E_F$, despite the lower variation of elemental substitution. Notably, the Cr-BFA sample exhibits a $\gamma$ pocket nearly 1.3 times larger than that of the BFA parent compound, while the Mn-BFA exhibits a $\gamma$ pocket slightly smaller than the parent compound, and therefore has the smallest $\gamma$ pocket among the studied compounds. 

This pronounced change in the $\gamma$ band for the Cr-BFA compound, compared to the minimal variation seen in the Mn-BFA compound, is consistent with a previous study on Cr- \citep{marli_Cr} and Mn-substitution \citep{marli_Mn}. In fact, the Mn-substitution study showed that even at higher nominal substitution levels, the $\gamma$ band undergoes little evolution relative to the pure compound. In contrast, the Cr-BFA sample exhibits a markedly enlarged $\gamma$ pocket, consistent with previous reports, indicating that Cr introduces hole doping and enhances the size of the hole pockets at the FS.

These results indicate that substitution has a stronger impact on the outer hole pocket in these compounds, with even minimal Cr substitution causing significant changes in the size of the $\gamma$ pocket. Moreover, they reinforce our previous findings for Cr- \citep{marli_Cr} and Mn-substitution \citep{marli_Mn}, which showed that Cr acts as an effective hole dopant, increasing the hole pockets near the Fermi level, while Mn does not introduce significant hole doping. Instead, the reduction in $T_\text{SDW}$ observed with Mn substitution is primarily attributed to enhanced electronic disorder and correlations, rather than charge doping.

\begin{figure*}[ht]
\centering
\includegraphics[width=\textwidth]{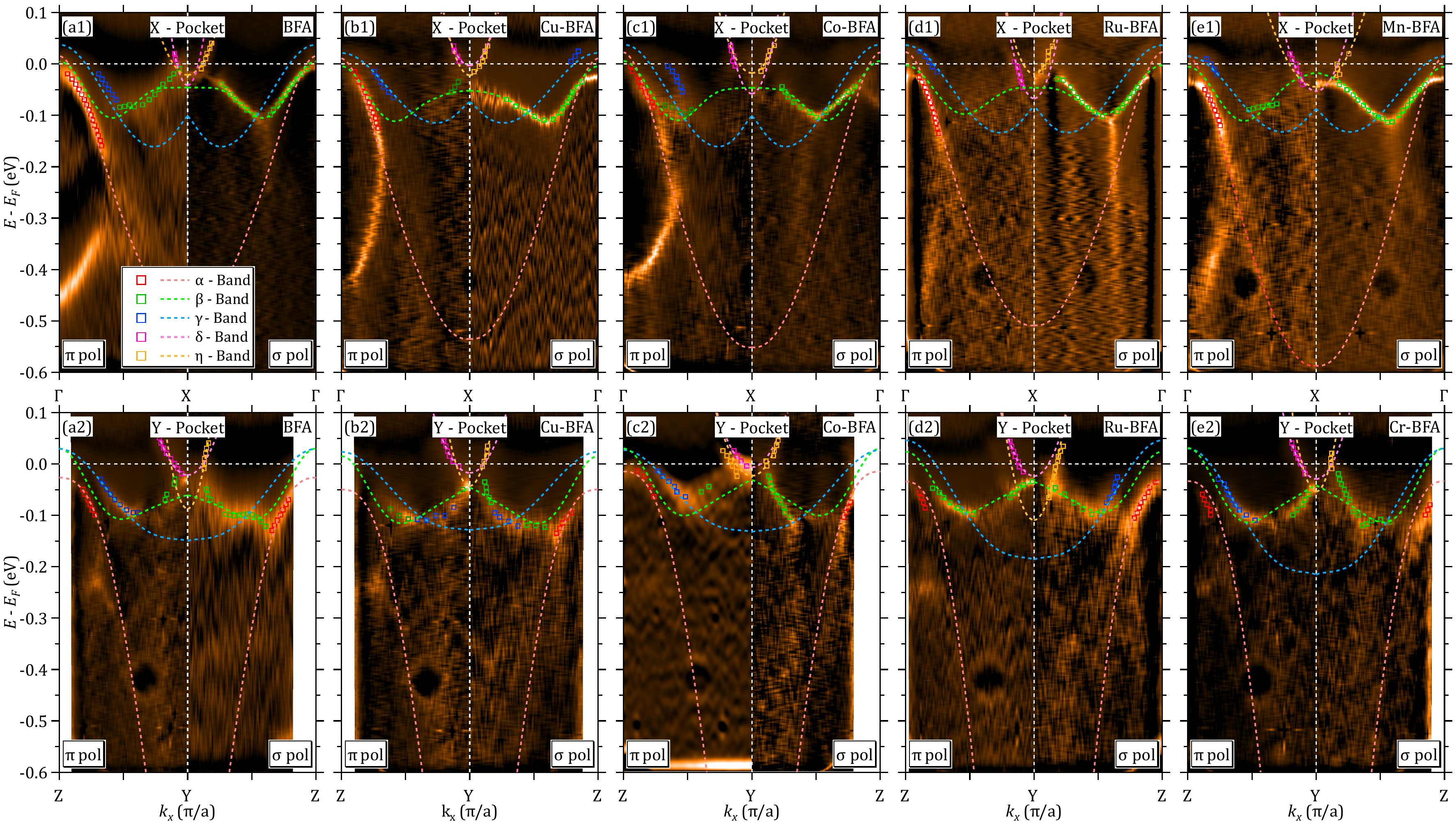}
\caption{Summary of the electron band dispersion analysis for the \FeFam{} family. Upper panels (a1 to e1) show data taken along the \GX{} direction focusing on the X pocket, while lower panels (a2 to e2) show data taken along the ZY direction focusing on the Y pocket. Both sets of data were measured using $\pi$ (left) and $\sigma$ (right) polarizations. For the special case of Mn-BFA and Cr-BFA, only the directions \GX{} (e1) and ZY (e2) for each compound were measured, respectively.\label{fig_all_XY_bands}}

\end{figure*}

Before analyzing the evolution of the hole band sizes as a function of macroscopic parameters, we first turn our attention to the electron pockets. As previously mentioned, a slightly different approach was used to extract the band positions for the electron pockets. In this case, we relied entirely on the band positions determined via the minima in the second derivative/curvature BMs. This alternative approach was necessary because fitting Lorentzian peaks to the MDCs proved extremely challenging due to the presence of multiple closely spaced and overlapping bands in the electron pockets. Moreover, only a small portion of the electron bands lies below $E_F$, so the majority remain undetectable by the ARPES technique.

To investigate the electron pockets, two specific momentum cuts were analyzed. As described in figure \ref{fig_opening}(d), Cut 1 was measured along the \GX{} direction, centered on the electron pocket, and Cut 3 was measured along the ZY direction, also centered on the electron pocket. Given that the compounds exhibit four-fold rotational symmetry in the paramagnetic state, the Y and X electron pockets are equivalent. This symmetry implies that measurements taken along these two directions effectively probe the same electronic structure, but from perpendicular directions.

Based on these considerations, we constructed figure \ref{fig_all_XY_bands}, which displays the second derivative or curvature BMs of the electron bands for all compounds studied. In these images, we used a gold color scale in which black represents maximum intensity and white corresponds to minimum intensity. The open squares indicate the intensity minima, which are interpreted as the band positions. We used the same color scheme adopted for the hole bands, red for $\alpha$, green for $\beta$, and blue for $\gamma$, and added two new colors to represent the electron bands: orange for the $\eta$ band and pink for the $\delta$ band. We define the $\delta$ band as the one typically observed with $\pi$ polarization (right side of each image), and the $\eta$ band as the one typically observed with $\sigma$ polarization (left side of each image).

The upper panels of figure \ref{fig_all_XY_bands} present the electron bands measured along the \GX{} direction, with $\pi$ polarization shown on the left and $\sigma$ polarization on the right. In the $\pi$ polarization data, a deeper electron band marked with pink squares is visible, which we assign to the $\delta$ band. Based on the orbital selection rules summarized in table \ref{tab_polari}, we attribute this band primarily to a \dxz{} orbital character. In the $\sigma$ polarization data, a shallow electron band marked with orange squares is identified as the $\eta$ band, which we associate primarily with a \dxy{} orbital contribution, one of the orbitals accessible in this polarization configuration. These observations are consistent with FS calculations of the BFA parent compound by Yin \textit{et al.}\citep{ba122_fs_the_yin}, which report along the \GX{} direction an outer electron pocket of \dxy{} character and an inner pocket of \dxz{} character. Similar conclusions were reached by Evtushinsky \textit{et al.}\citep{ba122_exp_fs_Evtushinsky} and Brouet \textit{et al.}~\citep{Brouet2012}, although they identify the shallow band as $\delta$ and the deeper one as $\eta$. Indeed, in our data the exact band bottoms are difficult to resolve, so we cannot definitively exclude the possibility that $\eta$ lies deeper than $\delta$. Nonetheless, since our primary goal is to compare the evolution of pocket size, which can be reliably inferred from our data, this ambiguity does not affect the conclusions.

In the measurements along the ZY direction (bottom panels of figure \ref{fig_all_XY_bands}), we were able to extract two electron bands from the Y pocket, along with hole bands from the Z pocket. Because the ZY direction is parallel to \GX{}, the same orbital selection rules from table \ref{tab_polari} apply. In the $\pi$ polarization, we identified a shallow electron band (the $\delta$ band, marked in pink), which we again attribute to a \dxz{} orbital character. In the $\sigma$ polarization, we identified a deeper electron band (the $\eta$ band, marked in orange), which we attribute to a \dxy{} orbital contribution. For the BFA parent compound, FS calculations show the $\eta$ band forms the outer pocket and the $\delta$ band forms the inner pocket, in agreement with the results of Yin \textit{et al.} \citep{ba122_fs_the_yin}.

Once the electron band positions were extracted, we fitted them with a parabolic model to generate the dashed lines shown in figure \ref{fig_all_XY_bands} (orange and pink dashed lines). For the hole bands, we used the same guide-to-eye models from figure \ref{fig_all_central_bands} (upper panels of figure \ref{fig_all_XY_bands}) or constructed new ones for the Z point hole bands (bottom panels of figure \ref{fig_all_XY_bands}).


The electron pocket measurements along the \GX{} direction for Cr-BFA and the ZY direction for Mn-BFA are not shown in figure \ref{fig_all_XY_bands}, as these data could not be acquired within the available beamtime. However, this does not affect the overall analysis and conclusions.

\begin{figure}[ht]
\centering
\includegraphics[width=\columnwidth]{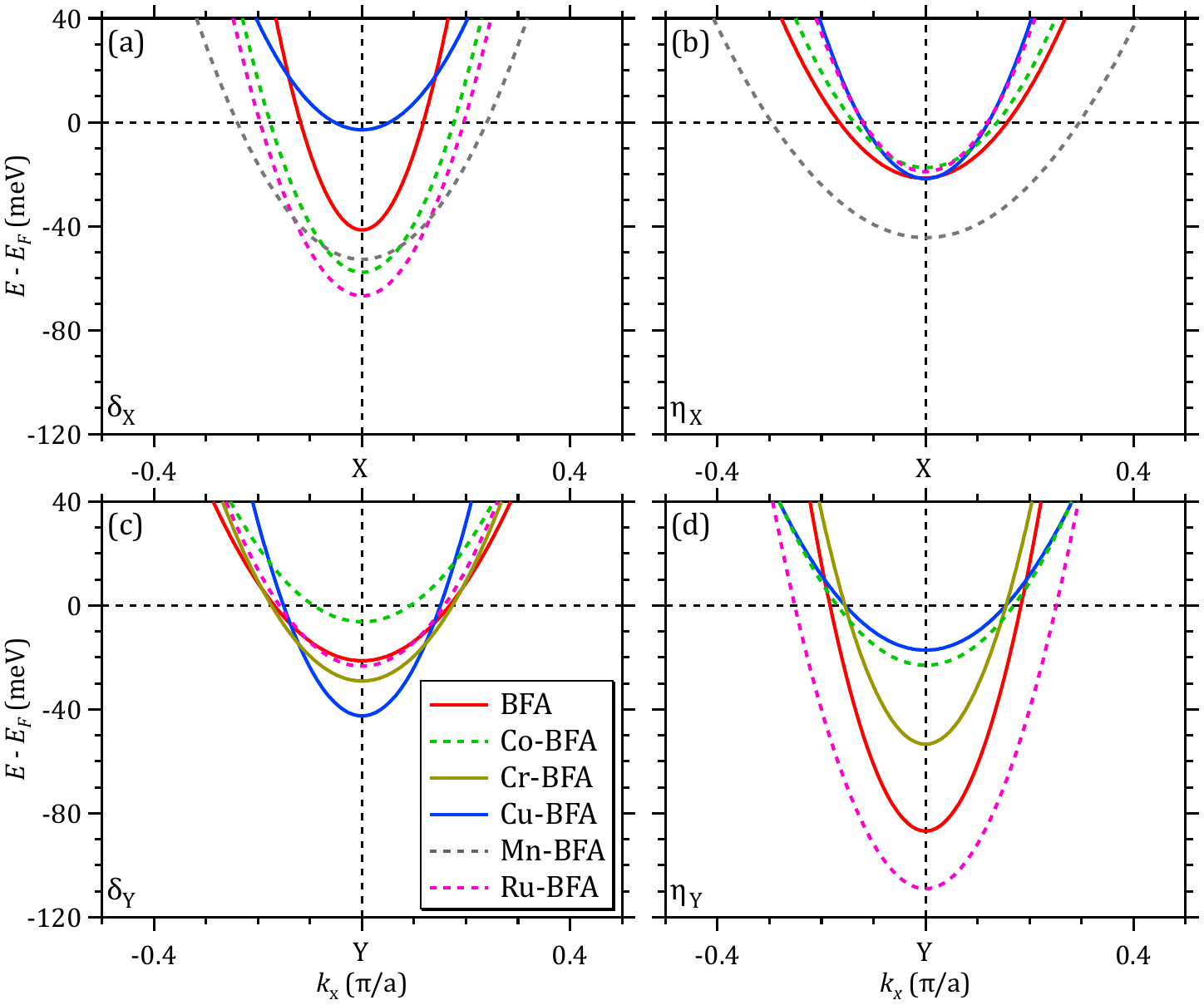}
\caption{Comparison of the electron bands across the \FeFam{} family. Upper panels (a and b) show the schematic dispersions of the two electron bands measured along the \GX{} direction in the electron pocket X: (a) $\delta_\text{X}$, and (b) $\eta_\text{X}$. Lower panels (c and d) present the same bands measured along the ZY direction in the electron pocket Y: (c) $\delta_\text{Y}$, and (d) $\eta_\text{Y}$. \label{fig_comp_e_pocket_bands}}

\end{figure}

Figure \ref{fig_comp_e_pocket_bands} summarizes the guide-to-eye fits for the electron bands. Panels (a) and (b) show the bands extracted from measurements along the \GX{} direction, while panels (c) and (d) correspond to measurements along the ZY direction.

It is important to emphasize that this analysis focuses on the variation in the size of the electron pockets, specifically the evolution of the Fermi wavevector $k_F$, rather than on the depth of the electron bands in binding energy, as the latter is more difficult to reliably extract from the data.

With this in mind, the plots clearly reveal substantial variations in electron pocket size across the different compounds. For instance, some compounds exhibit significantly larger electron pockets than the parent compound (\textit{e.g.}, Mn-BFA along the \GX{} direction), while others show reduced electron pocket sizes (\textit{e.g.}, Cu-BFA in both directions).

In the case of the Cr-BFA compound, a slight shrinking of the $\eta$ pocket can be observed, whereas the $\delta$ pocket remains essentially unchanged compared to the BFA parent compound. This observation is consistent with previous results on Cr-substituted BFA \citep{marli_Cr}, which reported an overall shrinking of the electron pockets along both the \GX{} and ZY directions, along with an expansion of the hole pockets. These trends confirm that Cr substitution acts as an effective hole dopant in the system.

On the other hand, for Mn-BFA we observe a pronounced enlargement of both electron pockets, at least along the \GX{} direction. Unfortunately, due to the absence of data along the ZY direction, it is not possible to directly determine whether the electron pockets expand or instead shrink in total area size. However, based on our earlier studies of Mn substitution in the BFA family \citep{marli_Mn}, we found a general trend of electron pocket shrinking, which suggests that despite the observed expansion along \GX{}, there may be a more significant contraction along ZY. This would result in an overall reduction of the electron pocket area. In either case, such changes would disrupt the nesting condition between electron and hole pockets, which is responsible for the antiferromagnetic ordering in these materials.

\begin{figure}[ht]
\centering
\includegraphics[width=\columnwidth]{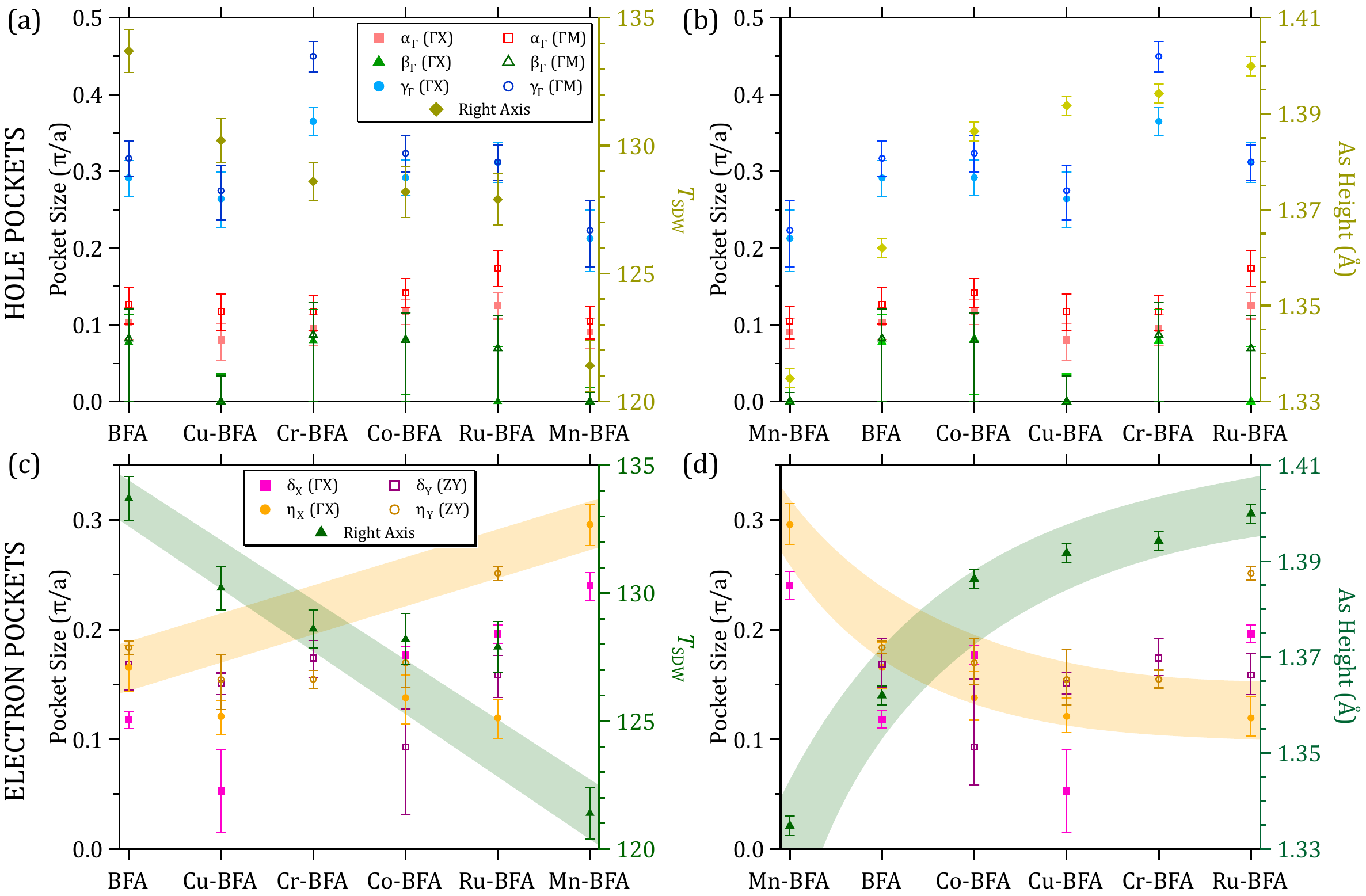}
\caption{Evolution of the hole (a and b) and electron (c and d) pocket sizes in the \FeFam{} family plotted as a function of the magnetic transition temperature $T_\textrm{SDW}$ (left panels) and As height with respect to the Fe plane (right panels). These graphs were constructed from the $k_F$ values of the bands in figures \ref{fig_comp_center_bands} and \ref{fig_comp_e_pocket_bands}. The right vertical axes in each panel correspond to the values of $T_\textrm{SDW}$ and As height, respectively.\label{fig_size_center}}

\end{figure}

Next, we examine the evolution of hole and electron pocket sizes in relation to certain macroscopic properties of the studied materials. Figures \ref{fig_size_center}(a) and (b) illustrate the evolution of the central hole pockets ($\alpha$, $\beta$, and $\gamma$), measured along the \GX{} direction (represented by filled circles and squares in light colors) and the \GM{} direction (open circles and squares in dark colors). Figures \ref{fig_size_center}(c) and (d), in turn, show the evolution of the electron pockets ($\delta$ and $\eta$), measured along the \GX{} direction (filled markers, light colors) and the ZY direction (open markers, dark colors). In the left-hand panels, the samples are sorted by their SDW magnetic transition temperature ($T_\mathrm{SDW}$), while in the right-hand panels, they are sorted by the As height relative to the Fe planes. The latter quantity was derived from Rietveld refinements of the X-ray diffraction data, enabling a direct determination of the lattice parameters, Fe–As bond length, and Fe–As–Fe bond angle.

As shown in Figure \ref{fig_size_center}(a), no clear correlation can be established between hole pocket sizes and $T_\mathrm{SDW}$. Even among doped samples with similar $T_\mathrm{SDW}$ values, the hole pocket sizes vary across compounds. Interestingly, for the $\gamma$ band, most doped samples exhibit larger pocket sizes compared to the BFA parent compound, with the notable exception of Mn-BFA. This suggests a general trend of increasing $\gamma$ pocket size upon transition-metal substitution at the Fe site. In contrast, the $\alpha$ and $\beta$ pockets remain relatively unchanged across different compounds, with variations falling within experimental uncertainty.

Similar conclusions can be drawn from figure \ref{fig_size_center}(b), where the same data are plotted as a function of the As height. No clear trend emerges between this structural parameter and the variation in hole pocket size.

Turning to the electron pockets shown in figures \ref{fig_size_center}(c) and (d), we observe in panel (c), where the samples are sorted by $T_\mathrm{SDW}$, a slight trend: the $\eta_X$ electron pocket (closed orange circles) tends to increase in size as $T_\mathrm{SDW}$ decreases (shaded area in green and orange). However, the most prominent correlation appears in panel (d), where the samples are sorted by As height. Here, a shaded region appears to show a clearer relationship: as the As height increases, the size of the $\eta_X$ electron pocket decreases. Since the $\eta_X$ band is primarily associated with the \dxy{} orbital, this observation suggests a link between structural distortions, suppression of $T_\mathrm{SDW}$, and modifications in the planar orbitals in the FS.

\begin{figure}[ht]
\centering
\includegraphics[width=\columnwidth]{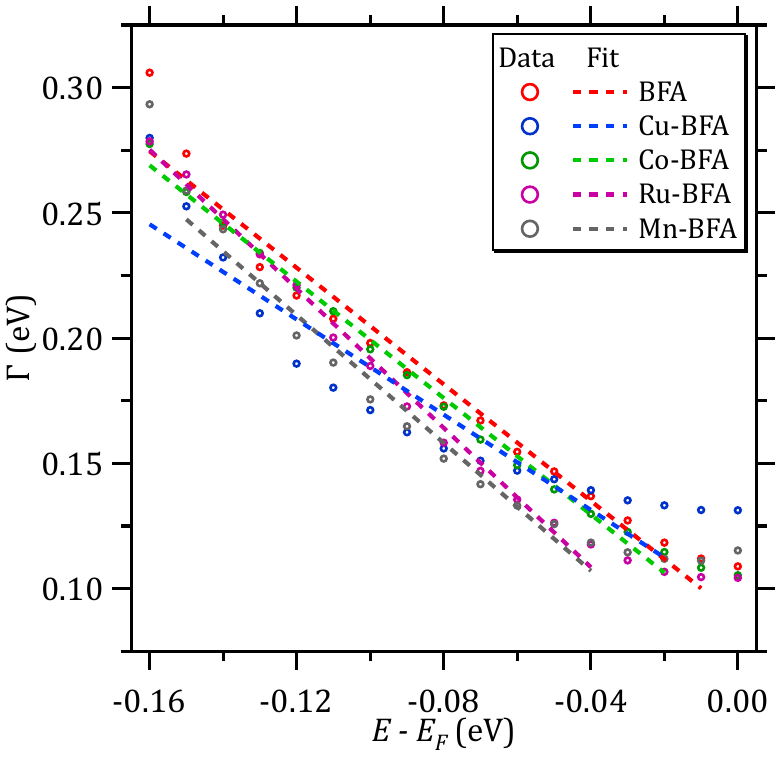}
\caption{Evolution of the scattering rate $\Gamma(E)$ across the \FeFam{} family as a function of binding energy $E$. The data were extracted from spectral fittings of the BMs measured along the \GM{} direction using $\sigma$ polarization.\label{fig_scatrate}}

\end{figure}

We now focus on the analysis of the spectral function itself. As previously mentioned, the ARPES spectral function depends on two key quantities: the renormalization function $Z(E)$ and the scattering rate $\Gamma(E)$. The latter, in particular, provides valuable insight into the underlying physics of these compounds and allows us to probe the degree of electronic correlations within the system.

For the experimental configuration along the \GM{} direction with $\sigma$ polarization, we were able to directly extract these quantities by fitting the MDCs using the spectral function described in Equation \ref{eq_arpes_spectra}. The extracted scattering rates for all compounds are shown in Figure \ref{fig_scatrate}. In most cases, $\Gamma(E_B)$ displays a linear dependence on the binding energy $E$, consistent with the behavior observed in our previous studies for Mn and Cr-substitute compounds\citep{marli_Mn, marli_Cr}. This linear trend contrasts with the quadratic dependence expected in a conventional Fermi liquid \citep{landau1959theory}, indicating that these materials exhibit strongly correlated metallic states, a hallmark of non-Fermi-liquid behavior \citep{Si2010,Stewart2001}, which can be described within the framework of marginal Fermi liquid theory \citep{Varma2020,Varma1989,fink_paper}. Indeed, a linear dependence of the scattering rate on energy or temperature has been widely reported in other correlated systems \citep{Varma2020,fink_paper}. 

It is important to note that data for the Cr-BFA compound are absent in this analysis. This omission is due to a misalignment of the crystal during measurement, which deviated from the intended \GM{} direction indicated in Figure \ref{fig_opening}(d). As a result, the data correspond to a region where our simple parabolic model for the $\beta$ band is no longer valid. To address this issue, we initially attempted to reconstruct a diagonal BM using the \GX{} FS measured with $\sigma$ polarization. While this approach provides a valid band dispersion for analyzing pocket sizes, it does not yield reliable results for extracting the scattering rate. Consequently, we excluded the Cr-BFA compound from this part of the analysis.

Nonetheless, in our previous study on Cr-substituted samples \citep{marli_Cr}, we found that the scattering rate for low Cr concentrations is comparable to that of the parent compound, exhibiting a slightly reduced slope parameter $b$.

\begin{table}[ht]
    \centering
    \caption{Summary of the scattering rate slope parameter $b$ extracted from linear fits, showing its evolution across the \FeFam{} family of compounds.}
    \label{tab_scatrate}
    \begin{tabular}{cccccc}
    \hhline{======}
        Compound & \hspace{0.01\textwidth}BFA\hspace{0.01\textwidth} & \hspace{0.01\textwidth}Cu-BFA\hspace{0.01\textwidth} & \hspace{0.01\textwidth}Co-BFA\hspace{0.01\textwidth} & \hspace{0.01\textwidth}Ru-BFA\hspace{0.01\textwidth} & \hspace{0.01\textwidth}Mn-BFA\hspace{0.01\textwidth}\\
        \hline
        $b$ & 1.16(6) & 0.95(10) & 1.16(3) & 1.39(3) & 1.28(7)\\
    \hhline{======}
    \end{tabular}
\end{table}

Furthermore, we successfully fit the scattering rate data using a linear model $\Gamma(E)=\Gamma_0+bE$, obtaining a typical slope value of $b \approx 1.1$ for most compounds, consistent with the results reported by Fink \textit{et al.} \citep{fink_paper} for low-substituted systems (see Table \ref{tab_scatrate}). In that study, a correlation was suggested between the emergence of superconductivity and the value of $b$: materials near the underdoped and overdoped regions of the phase diagram tend to have $b \approx 1$, whereas optimally doped compounds exhibit $b$ values above 3. Notably, all the compounds investigated in our study lie within the underdoped regime and show consistent results with those reported by Fink \textit{et al.} \citep{fink_paper}.

\section{Summary and Conclusions}

We have identified a clear evolution of the outer hole pocket ($\gamma$) as a function of elemental substitution, despite all samples exhibiting similar SDW transition temperatures. Notably, Cr-substituted BFA displays a $\gamma$ pocket approximately 1.3 times larger than that of the parent BFA compound, while Mn-substituted BFA shows the smallest $\gamma$ pocket among all samples studied. This observation aligns with our previous work on Mn substitution \citep{marli_Mn}, which demonstrated minimal modifications to the parent band structure. In contrast, our earlier study on Cr substitution \citep{marli_Cr} revealed more significant changes in the outer hole band, highlighting a sharp contrast between the effects of Mn and Cr.


In the low-substitution regime, we find no clear correlation between the size of the central hole pockets ($\alpha$ and $\beta$) and either $T_\text{SDW}$, the As height, or the electron/hole character of the substitution. In contrast, more systematic changes are observed in the Fermi surface pockets derived from planar orbitals (\dxy{} and \dxxyy{}), including the outer hole pocket ($\gamma$) and the electron pockets ($\delta$ and $\eta$). In particular, the $\eta$ pocket shows a moderate enlargement correlated with decreasing $T_\text{SDW}$, as shown in Figure \ref{fig_size_center}(c). These observations, combined with previous EXAFS studies \citep{bittar_exafs} linking substitution to a reduced Fe–As bond length, support the interpretation that bands with planar orbital character are especially sensitive to local structural distortions and are central to the evolution of magnetism and superconductivity in lightly substituted \FeFam{}.

To test this hypothesis, we compared the size of the planar electron pocket with the As height extracted from our x-ray diffraction measurements. While x-ray diffraction is not the most precise method for determining the As height within the FeAs tetrahedra, we nevertheless find a consistent correlation: as the As height increases, the planar electron pocket size decreases.

Additionally, we observed a predominantly linear dependence of the scattering rate on binding energy across most compounds studied. This behavior is indicative of strong electronic correlations and provides further evidence for marginal Fermi-liquid behavior in these materials.


\section{Acknowledgments}

We acknowledge the Canadian Light Source (CLS) for time on Quantum Materials Spectroscopy Centre (QMSC) beamline  under Proposal No. 014139 and support of Sergey Gorovikov. K.R.P acknowledge funding from the CNPq (grant 162327/2017-0) and CAPES (Finance Code 001); K.R.P., A.P.M., M.M.P., G.S.F., H.B.P., F.A.G., P.G.P. and C.A. acknowledge financial support from FAPESP (grant 2017/10581-1, 2023/17024-1, 2021/02763-8, 2018/08040-5 and 2024/13291-8) and CNPq (Grants No. 405408/2023-4 and 311783/2021-0).  We acknowledge the support of the INCT project Advanced Quantum Materials, involving the Brazilian agencies CNPq (Grant 408766/2024-7), FAPESP, and CAPES.

\bibliographystyle{apsrev4-2}
\bibliography{bibliography}

\end{document}